\documentclass[aps,nofootinbib]{revtex4} 
\usepackage{graphicx}
\input epsf

\def\cmm2{{\,\rm cm^{-2}}}
\def\cm2{{\,{\rm cm}^2}}
\def\cmm3{{\,{\rm cm}^{-3}}}
\def\gcmm3{{\,{\rm g\,cm^{-3}}}}

\def\fun#1#2{\lower3.6pt\vbox{\baselineskip0pt\lineskip.9pt
  \ialign{$\mathsurround=0pt#1\hfil##\hfil$\crcr#2\crcr\sim\crcr}}}

\def\be{\begin{equation}}
\def\ee{\end{equation}}
\def\bea{\begin{eqnarray}}
\def\eea{\end{eqnarray}}
\newcommand{\vs}{\nonumber\\}

\newcommand{\ec}[1]{Eq.~(\ref{eq:#1})}
\newcommand{\Ec}[1]{(\ref{eq:#1})}
\newcommand{\eql}[1]{\label{eq:#1}}

\begin{document}

\title{Second Order Geodesic Corrections to Cosmic Shear}

\author{S. Dodelson$^{1,2,3}$, E.~W.~Kolb$^{1,2}$, S.~Matarrese$^{4,5}$, A.~Riotto$^{4}$ and P. Zhang$^1$}

\affiliation{$^1$NASA/Fermilab Astrophysics Center Fermi National
Accelerator Laboratory, Batavia, IL~~60510-0500}
\affiliation{$^2$Department of Astronomy \& Astrophysics, The
University of Chicago, Chicago, IL~~60637-1433}
\affiliation{$^3$Department of Physics and Astronomy, Northwestern
University, Evanston, IL~~60208} 
\affiliation{$^4$Dipartimento di Fisica ``G. Galilei,'' Universita
di Padova, Italy}
\affiliation{$^5$INFN, Sezione di
Padova, via Marzolo 8, I-35131, Italy}

\date{\today}
\begin{abstract}
We consider the impact of second order corrections to the geodesic equation
governing gravitational lensing. We start from the full
second order metric, including scalar, vector and tensor perturbations, 
and retain all relevant contributions to the cosmic shear corrections 
that are second order in the gravitational potential.
The relevant terms are: the nonlinear evolution of the scalar gravitational potential,
the Born correction, and lens-lens coupling. No other second order terms 
contribute appreciably to the lensing signal. Since ray-tracing
algorithims currently include these three effects, this derivation 
serves as rigorous justification for the numerical predictions.
\end{abstract}
\maketitle

\section{Introduction}

Gravitational lensing of background galaxies by large scale structure offers an
excellent way to study the distribution of matter in the 
universe\,\cite{Kaiser:1991qi,Mellier:1998pk,Bartelmann:1999yn,Dodelson}. 
Measurements of the cosmic shear
are already enabling us to constrain the dark matter
abundance\,\cite{Hoekstra:2002nf,VanWaerbeke:2004af}. In the future,
large surveys may well uncover properties of dark energy, such as its 
abundance and equation of 
state\,\cite{Benabed:2001dm,Refregier:2003xe,Takada:2003ef,Heavens:2003jx} 
and of
neutrinos\,\cite{Abazajian:2002ck}. This program will be successful only if
we can make very accurate predictions about what theories 
predict\,\cite{Huterer:2004tr}.

Predicting the cosmic shear signal is more difficult than making predictions
for the cosmic microwave background but much more straightforward than
for the galaxy distribution.
The former rely predominantly on linear perturbation theory so are extremely
robust. The latter require not only nonlinear evolution but also an
understanding of how the galaxy distribution is related to the mass. For
lensing, we need to account for nonlinearities, i.e. move beyond 
linear perturbation theory, but, since the light
trajectories depend only on the mass, we do not need to worry about how and
where galaxies form\footnote{We refer here to cosmic shear. 
Other lensing phenomena, such as
galaxy-galaxy lensing, do involve the interplay of
light and mass.}. 

The obvious way to treat nonlinearities is to perform numerical simulations of
structure formation. Since dark matter is much more abundant than baryonic
gas, only N-Body simulations are needed, not\footnote{This ceases to be true on
very small scales where the effects of baryons can no longer 
be neglected\,\cite{White:2004kv,Zhan:2004wq}.} the much more computational costly 
hydrodynamical simulations. A number of groups have run such simulations and
extracted predictions by {\it ray-tracing} photon paths through the simulated
universe\,\cite{jainseljak,White:2003xz}. In principle, this 
proceedure accounts for some 
second order effects, such as the
second order perturbations to the potential and the fact that the path of the
photons is not simply a straight line. However, the ray-tracing algorithims use
the first-order geodesic equation to find the distortions in each slice of the
universe. Therefore, again in principle, the algorithms may be missing
relevant terms.
A fully consistent second order calculation is
warranted.

In this paper, we start from the second
order metric, proceed to obtain the second order geodesic equation, and then
compute the cosmic shear consistently retaining all second order terms. 
The rest of the paper is organized as follows: \S{II} reviews the first order
calculation, and \S{III} presents an order of magnitude estimate that
will prove useful when wading through a host of second order terms. 
\S{IV} presents the full second order
geodesic equation, borrowing heavily from previous work. In it, we winnow
out the
terms that contribute negligibly, keeping only those terms that are
responsible for observable deflections. 

\section{First Order Shear}

Consider the first order deflection of light due to inhomogeneities along the line of
sight. We work in a flat universe; then the 
geodesic equation may be integrated to give 
 \be x_\perp^{(1)i}(\vec\theta) =
 \int_{0}^{\chi_s} d\chi \, (\chi_s-\chi) f_\perp^{(1)i}(\chi,\vec\theta)
.\eql{xone}\ee 
Here $x_\perp$ is the perpendicular deflection of a ray starting at 
comoving distance away $\chi_s$ and traveling towards us at $\chi=0$ detected at angular position
$\vec\theta$ with respect to a fixed $z$-axis.
The first superscript on $x_\perp$ denotes the order of the perturbation,
the second is a space-time index; we are interested only in the two components
perpendicular to the line of sight. The first order distortion 
is
\be
f_\perp^{(1)i}(\chi,\vec\theta) = -\Gamma^{(1)i}_{\alpha\beta} p^{(0)\alpha} p^{(0)\beta} 
.\eql{fone}\ee
Here $\Gamma^{(1)}$ is the first order Christoffel symbol and
the zero-order space-time vector
$p^{(0)\alpha}=(1,-e^i)\simeq (1,-\theta^i,-1)$, where $e^i$ is
the spatial direction vector. The square of $\hat e$ is unity, and
$\theta^i$ is the transverse displacement from the $z$-axis. We work
in the small angle approximation throughout, so
$\theta$ is assumed small, and the $z$-component of $\hat e$ is approximately
unity. 

In the contraction of the Christoffel symbol with the two factors
of $p$, the only terms which contribute are those with
$\alpha,\beta=0$ or $3$. 
Using standard results\,\cite{Dodelson}, 
we can write the contraction as \be
\Gamma^{(1)i}_{\alpha\beta} p^{(0)\alpha} p^{(0)\beta} =
2{\partial\phi^{(1)}(\vec x^{(0)}(\theta,\chi))\over\partial x^i}
\eql{gamone}\ee 
where $\phi^{(1)}$ is the first-order 
gravitational potential evaluated at 
the unperturbed\footnote{Evaluating it
at the perturbed position leads to a second order term, 
and for now we are considering only first
order terms.} position
of the light: $\vec x^{(0)}(\theta,\chi)=\chi[\vec\theta,1]$. 
So, \be x_\perp^{(1)i} =
-2\int_{0}^{\chi_s} d\chi\, (\chi_s-\chi)
{\partial\phi^{(1)}(\vec x^{(0)}(\theta,\chi))\over\partial x^i}\eql{xperp}\ee 

The shear matrix is defined as the derivative of this perpendicular deflection
with respect to observed angle $\vec\theta$:
\be
\psi_{ij}(\vec\theta,\chi_s) \equiv  
{1\over \chi_s} {\partial x_\perp^i\over \partial \theta^j}
.\eql{defshear}\ee
To lowest order of course $x_\perp^{(0)i}=\chi_s\theta^i$, 
so the zeroth order term in the shear
matrix is simply the identity. From \ec{xperp}, we see that the 
first order term in the
shear matrix is
\bea
\psi^{(1)}_{ij}(\vec\theta,\chi_s) &=& -2\int_0^{\chi_s} d\chi\, {\chi_s-\chi\over\chi_s}
{\partial^2\phi^{(1)}(\vec x^{(0)}(\theta,\chi))\over\partial x^i \partial \theta^j}
\vs
&=&
-2\int_0^{\chi_s} d\chi\, \chi {\chi_s-\chi\over\chi_s}
{\partial^2\phi^{(1)}(\vec x^{(0)}(\theta,\chi))\over\partial x^i \partial x^j}
.\eql{shear}\eea
Again the second equality here follows because the potential 
is evaluated along the unperturbed
path of the light.

\section{Order of Magnitude Estimate}

In order to wade through the second order terms and extract the most relevant
ones, we need a way of doing order of magnitude estimates. From \ec{shear}, the
shear induced by a perturbation with wavelength $\lambda$ is of order
$(r_H/\lambda)^2\phi$ since cosmological distances $\chi$ are of order 
the Hubble radius, $r_H=3000$h$^{-1}$ Mpc. The mean of the shear of course
vanishes, but its rms we would expect to be of order 
$(r_H/\lambda_{\rm max})^2\phi_{\rm rms}$, where
$\lambda_{\rm max}$ is the wavelength near which
perturbations contribute the most to the
deflections. In currently popular models, 
$\lambda_{\rm max} \sim 30$h$^{-1}$ Mpc. 
The
rms amplitude of the gravitational potential is of order $10^{-5}$, so one
would naively expect the rms shear to be about ten percent. 
The rms is actually 
smaller than this because not all Fourier modes contribute to the variance.
Perturbations which vary rapidly in the $z$-direction lead to 
little total distortion since regions of positive and negative overdensity along
the line of sight cancel each other. Thus, in the 3D Fourier space only modes 
with very small $k_3$ ($\le r_H^{-1}$) contribute. The fraction of modes 
which satisfy this constraint is $(\lambda/r_H)$ for a given
$\lambda$\,\cite{Kaiser:1991qi,Dodelson}; 
the variance is therefore
smaller than the naive estimate by this fraction, and the rms by 
its square root. The
rms amplitude is therefore of order 
$(r_H/\lambda_{\rm max})^{3/2} \phi_{\rm rms}$,
less than a percent.

There are two lessons we learn from this order of magnitude estimate which will
be useful when we evaluate second order terms. First, we need consider only those
terms in $f^{(2)}$ which vary little along the line of sight. Terms with partial
derivatives with respect to $z\equiv x^3$ can be neglected. Second,
the shear is much larger than the depth of the typical potential 
well, $\phi_{\rm rms}$.
It is the rapid changes in the potential which lead to large deflection; i.e.,
$(\partial\phi/\partial x^i) r_H \gg \phi$ (as long the derivatives
are in one of the transverse directions, $i=1,2$). When considering second order terms,
therefore, we will be most impressed by those with the most derivatives.

Armed with this information, let's return to \ec{gamone} for $f^{(1)i}_\perp$
and consider the correction incurred when we account for the fact that the
photon does not travel along a straight line so the argument of $\phi$ might
reasonably be taken as $\vec x^{(0)} + \vec x^{(1)}$. Expanding about the zero
order path then leads to one possible second order term
\be
{\partial^2\phi\over \partial x_i \partial x_j}\Big\vert_{x=x^{(0)}}
x^{(1)j} 
\eql{born}\ee
This term is known as the {\it Born correction}\cite{Bernardeau:1996un,Schneider:1997ge}. We will soon see that it emerges
as one of many second order terms. To derive it here, we have cheated since the
argument of $\phi$ in \ec{gamone} is the {\it zero order} path. We 
introduce this term now only because we know it will show up in the second order
zoo, and we want to estimate its order of magnitude. From \ec{xperp}, $x^{(1)}$ is of
order $r_H^2\partial\phi$, so the Born correction is of order
$(\partial^2\phi) r_H^2 \partial \phi$.
Thus, we need keep second
order terms in $f_\perp^{(2)}$ only if they are of order 
$r_H^2\partial^3\phi^2$.

\section{Second Order Geodesic Corrections}

The perpendicular deflection to second order is given by the analogue of
\ec{xone}, with the superscripts changed from $^{(1)}$ to $^{(2)}$,
and\,\cite{cp95}
 \bea
f^{(2)i}_\perp&=&-\Gamma^{(0)i}{}_{\alpha\beta}p^{(1)\alpha}p^{(1)\beta}
  -2\Gamma^{(1)i}{}_{\alpha\beta}p^{(0)\alpha}p^{(1)\beta}
 -2\partial_\sigma\Gamma^{(0)i}{}_{\alpha\beta}
  x^{(1)\sigma}p^{(0)\alpha}p^{(1)\beta} 
 \vs &-&\partial_\sigma\Gamma^{(1)i}{}_{\alpha\beta}
  x^{(1)\sigma}p^{(0)\alpha}p^{(0)\beta}
   -{1\over 2}
  \partial_\sigma\partial_\tau\Gamma^{(0)i}{}_{\alpha\beta}
  x^{(1)\sigma}x^{(1)\tau}p^{(0)\alpha}p^{(0)\beta} 
 -
  \Gamma^{(2)i}{}_{\alpha\beta}p^{(0)\alpha}p^{(0)\beta}
  .\eql{carroll}
  \eea
 The zero order and first order
 Christoffel symbols are well-known~\cite{kt,Dodelson}, while the second
 order $\Gamma^{(2)}$ has been computed by Bartolo et
 al.\,\cite{Bartolo:2004if}. The zero order direction vector $p^{(0)}$ is given
 after \ec{fone}, while the time and space components of the
 first order direction vector are\,\cite{cp95}
 \bea p^{(1)0}(\chi)
& =& p^{(1)0}(\chi=0) - \int_{0}^\chi
  f^{(1)0}(\chi')d\chi' \ ,\vs
  p^{(1)i}(\chi) & = &- \int_{0}^\chi
  f^{(1)i}(\chi') d\chi' \ \eql{direction}\eea
 with $f^{(1)\alpha} = -\Gamma^{(1)\alpha}{}_{\mu\nu} p^{(0)\mu} p^{(0)\nu}$.
To compute the first and second order Christoffel symbols, we need to specify
the metric, thereby choosing a gauge. 
Following the formalism of
\cite{Bartolo:2004if}, we set
\bea
g_{00} &=& -a^2\left[ 1 + 2\phi^{(1)} + \phi^{(2)} \right]
\vs
g_{0i} &=& a^2\left[ {1\over 2} \partial_i\omega^{(2)}
+ {1\over 2} \omega^{(2)}_i \right]
\vs
g_{ij} &=& a^2 \Big[ \left( 1-2\psi^{(1)} -\psi^{(2)} \right) \delta_{ij}
+{1\over 2} D_{ij}\chi^{(2)}  
+ {1\over 2} \left( \partial_i\chi^{(2)}_j + \partial_j\chi^{(2)}_i
+\chi^{(2)}_{ij} \right) \Big].
\eql{metric}\eea
Scalar perturbations are described by $\phi$, $\psi$, $\chi$ and $\omega$; 
vector perturbations by $\omega_i^{(2)}$ and $\chi^{(2)}_i$ 
; and tensors by $\chi^{(2)}_{ij}$. Finally
\be
D_{ij} \equiv \partial_i\partial_j - {\delta_{ij} \over
3} \nabla^2
 .\ee
Note that we assume that there are no first order vector or tensor
perturbations. Even without these, scalar, vectors, and tensors mix at second
order, so $\chi^{(2)}_{ij}$ and $\omega^{(2)}_i$ and  $\chi^{(2)}_i$ 
are generally non-zero, quadratic in first order
scalar perturbations.

We now work through the terms in \ec{carroll} explicitly, retaining only those
of order $r_H^2 \partial^3\phi^2$ or higher.

\begin{itemize}

\item{First Term: $-\Gamma^{(0)i}{}_{\alpha\beta}p^{(1)\alpha}p^{(1)\beta}$} 

The zero order
Christoffel symbol is proportional to the Hubble rate and is
non-zero only when one of the lower indices is equal to $i$ and
the other equal to zero. Thus this first term reduces to \be
-\Gamma^{(0)i}{}_{\alpha\beta}p^{(1)\alpha}p^{(1)\beta} =
-2r_H^{-1}p^{(1)0}p^{(1)i}.\ee
Forgetting about the boundary term in the first of \ec{direction}, and remembering
that $\chi\sim r_H$, we see that $p^{(1)0}\sim r_H f^{(1)0}$ and 
$p^{(1)i}\sim r_Hf^{(1)i}$. So this first term in $f^{(2)i}$ 
is of order $r_H f^{(1)0} f^{(1)i}$. We showed in \S{II}
that $f^{(1)i}=2\phi_{,i}$ so this first term is of order $r_H\partial\phi
f^{(1)0}$; it contributes appreciably only if $f^{(1)0}$ is of order 
$r_H\partial^2\phi$ or higher.
But
 \be
f^{(1)0}=-\Gamma^{(1)0}_{\alpha\beta}p^{(0)\alpha}p^{(0)\beta}
.\ee 
Recall that $p^{(0)\alpha}$ is negligible unless $\alpha=0$
or 3. If both $\alpha$ and $\beta$ are zero, then the Christoffel
symbol is $\dot\phi\sim r_H^{-1}\phi\ll r_H \partial^2\phi$. If both
indices are equal to 3, the Christoffel symbol again is of order
$\phi/r_H$, i.e. negligible. If one of the
indices is zero and the other equal to three, the Christoffel
symbol is equal to $\partial_3\phi$, which is negligible for all perturbations
of interest, i.e. for all perturbations which vary little along the line of
sight. Thus this first term does not contribute.

\item{Second Term: $-2\Gamma^{(1)i}_{\alpha\beta}p^{(0)\alpha}p^{(1)\beta}$} 

Since only the $\alpha=0$ or 3 components of $p^{(0)}$ are non-negligible,
\be
 -2\Gamma^{(1)i}_{\alpha\beta}p^{(0)\alpha}p^{(1)\beta}
\rightarrow
-2\Gamma^{(1)i}_{0\beta}p^{(1)\beta} + 2\Gamma^{(1)i}_{3\beta}p^{(1)\beta}
\eql{termt}\ee
Recall from the previous paragraph that 
\bea
p^{(1)0} &\sim &\phi
\vs
p^{(1)i} &\sim& r_H\partial_i\phi
\eea
So the biggest contribution will come from $\beta=j$, with $j=1$ or 2, one of
the transverse directions. The contribution is
of order $r_H\partial\phi\Gamma$. But 
$\Gamma^{(1)i}_{0j}=-\dot\psi\delta_{ij}\sim
r_H^{-1}\phi$, so the first term on the right in \ec{termt}
is of order $\partial\phi^2$ and
can be neglected. The second term in \ec{termt},
$2\Gamma^{(1)i}_{3j}p^{(1)j}$, is of order
$r_H\partial\phi\Gamma^{(1)i}_{3j}$ ; recall~\cite{Dodelson} that this 
Christoffel
symbol sets rotating pairs of its indices equal to each other, with the other
index applying to the derivative of the potential. We certainly then do not want
the index $_3$ to be the derivative since $\partial_3$ is very small. 
But both $i$ and $j$ are transverse indices so cannot be equal to 3:
all terms here vanish.

\item{Third Term: $-2\partial_\sigma\Gamma^{(0)i}{}_{\alpha\beta}
  x^{(1)\sigma}p^{(0)\alpha}p^{(1)\beta}$} 
  
After invoking some of the
approximations used in the previous paragraphs, we get
\be
-2\partial_\sigma\Gamma^{(0)i}{}_{\alpha\beta}
  x^{(1)\sigma}p^{(0)\alpha}p^{(1)\beta}
\rightarrow
-2\partial_\sigma\Gamma^{(0)i}{}_{0j}
  x^{(1)\sigma} p^{(1)j}.
\ee
The derivative here acts only on the Christoffel symbol, which depends only
on time. Thus $\sigma=0$, and $\partial\Gamma\sim r_H^{-2}$. 
The first order perturbation in
$x^{(1)0}$ is of order $r_H \phi$, so this term is of order $\partial\phi^2$, far too small
to contribute.

\item{Fourth Term: $-\partial_\sigma\Gamma^{(1)i}{}_{\alpha\beta}
  x^{(1)\sigma}p^{(0)\alpha}p^{(0)\beta}$} 
  
This is the only term in \ec{carroll} quadratic in the
first order variables (i.e., the only one of the first five terms) which
contributes. It gives the Born correction and lens-lens coupling, 
a very nice result emerging naturally from this second
order formalism. To see this, note that the only 
terms in the implicit sum which contribute are
those with $\alpha=\beta=0$ or $\alpha=\beta=3$. 
Both of these contribute identically, so this term is
\be
-2\left[ \partial_\sigma (\partial_i \phi) \right] x^{(1)\sigma}
= -4 {\partial^2\phi\over \partial x^i\partial x^j} \int_{0}^\chi d\chi' 
(\chi-\chi')
{\partial \phi\over \partial x^j}
.\ee

\item{Fifth Term: $-(1/2)
  \partial_\sigma\partial_\tau\Gamma^{(0)i}{}_{\alpha\beta}
  x^{(1)\sigma}x^{(1)\tau}p^{(0)\alpha}p^{(0)\beta}$}
  
  This term vanishes since the zero order direction vectors are non-negligible
  only if the indices $\alpha$ and $\beta$ are set to 0 or 3, and
  $\Gamma^{(0)i}{}_{00}=\Gamma^{(0)i}{}_{03}=\Gamma^{(0)i}{}_{33}=0$.

\item{Sixth Term: $-\Gamma^{(2)i}{}_{\alpha\beta}p^{(0)\alpha}p^{(0)\beta}$}

The only remaining term is the one explicitly second order in $\Gamma$. Before delving into
this term, we can localize it further by noticing that the indices $\alpha$ and $\beta$ must be either 0 or 3. Thus, this last term is
\be
- \left[ \Gamma^{(2)i}{}_{00}+\Gamma^{(2)i}{}_{33}-2\Gamma^{(2)i}{}_{03}  \right]
.\ee

To compute $\Gamma^{(2)}$, we lift the results from the Appendix 
of Ref.~\cite{Bartolo:2004if}. There, they compute the
Christoffel symbols to second order. There are a number of 
simplifications we can make. First, we work in a gauge (the so-called 
Poisson gauge) in which the scalars $\omega^{(2)}$ and $\chi^{(2)}$ 
as well as the vector  $\chi^{(2)}_i$ are set to zero, so  
vector perturbations are described only by $\omega^{(2)i}$, 
henceforth simply called $\omega$. 
Similarly, we assume there are no tensor perturbations at first order, 
and the second order perturbations are described solely by
the traceless and divergenceless tensor $\chi_{ij}$ 
[spatial indices are raised and lowered with the Euclidean metric, so 
we take no care in distinguishing upper from lower indices]. 
Within this framework,
\be
\Gamma^{(2)i}_{00} + \Gamma^{(2)i}_{33} -2 \Gamma^{(2)i}_{03} =
{1\over 2} \partial^i\left[ \phi^{(2)} + \psi^{(2)} \right] 
+ 4\phi^{(1)} \partial^i\left[ \phi^{(1)} \right]
+
{1\over 2} \left[ \omega^{i}{}' - \partial_3 \omega^i + \partial_i \omega^3 \right]
+ {1\over 4} \left[ 2\partial_3\chi_{i3} - \partial_i \chi_{33} - 2\chi_{i3}'\right]
.\eql{eso}\ee
Here prime denotes differentiation with respect to conformal time. 

To weigh the relative importance of each of these terms, it is important to remember that the only terms that
contribute are those which are of order $r_H^2\partial^3\phi^2$. 
This results in both vector and tensor modes being irrelevant.
Let's work through the vector case explicitly; the tensor case is similar. 
The vector modes have been
worked out in Ref.~\cite{Mollerach:2003nq}. Eqs. (17-19) there give explicit 
expressions for the second order vector
perturbations (their $V=-\omega/2$). By counting powers of 
$k\rightarrow\partial$, we see
that $\omega^i\sim r_H\partial\phi^2$. Even if we forget about the 
constraint that $\partial_3$ must be small, this means that
the second line is at most of order $r_H\partial^2\phi^2$, too small 
by a factor of order $r_H/\lambda_{\rm max}$ to contribute. A similar argument
applied to the expression for tensor modes in Ref.~\cite{Mollerach:2003nq} 
leads to a similar conclusion. Thus, the only
relevant explicit second order terms are from scalar perturbations, 
those on the first line of \ec{eso}. Using the
same order of magnitude estimate, we can immediately eliminate the 
second scalar term on the first line, since it is
of order $\partial\phi^2$, clearly too small to be relevant. 

The only relevant explicit second order terms therefore are
\be
\Gamma^{(2)i}_{00} + \Gamma^{(2)i}_{33} -2 \Gamma^{(2)i}_{03} =
{1\over 2} \partial^i\left[ \phi^{(2)} + \psi^{(2)} \right]
\rightarrow \phi^{(2)}_{,i}.
\ee
The final limit comes from recognizing that anisotropic stress, inducing a
$\psi-\phi \neq 0$ term, is a post-Newtonian second-order contribution; this 
can only be relevant either on very large scales or on very small highly 
non-linear scales \cite{carbmat}, 
so $\psi=\phi$ to a good approximation for the range 
of wavelenghts of interest here.  

\end{itemize}

Collecting second order
terms and adding to the first order term, we emerge with an expression
for the perpendicular deflection accurate to second order in the gravitational
potential:
\begin{widetext}
\bea
x_\perp^i(\vec\theta) &=& -2 \int_0^{\chi_s} d\chi (\chi_s-\chi) \Bigg[
{\partial \over\partial x^i} \Big( 
\phi^{(1)}\big(\vec x^{(0)}(\chi,\vec\theta)\big) 
+ {1\over 2} \phi^{(2)}\big(\vec x^{(0)}(\chi,\vec\theta)\big)
\Big) \vs
&& \qquad\qquad+ 2 {\partial^2\phi^{(1)}\big(\vec x^{(0)}(\chi,\vec\theta)\big)
 \over\partial x^i\partial x^k} 
 \int_0^\chi d\chi' (\chi-\chi') 
 {\partial\phi^{(1)}\big(\vec x^{(0)}(\chi',\vec\theta)\big)\over\partial x^k}
 \Bigg].
\eql{xtwo} \eea
\end{widetext}
Note that the first set of terms on the right $\phi^{(1)}+\phi^{(2)}/2$
is simply the full nonlinear potential out to second order; i.e., 
the metric (\ec{metric}) contains this combination. So we might expect 
the full deflection to be sensitive to the fully nonlinear gravitational
potential. Indeed this is the assumption built into all previous work. Our second
order treatment justifies this assumption. Also note that the second term is
indeed the Born correction alluded to in \ec{born} since twice the inner integral
is equal to $x^{(1)}$. In fact, the so-called lens-lens correction
is also included in this term: when differentiating $x_\perp^i$ to get the shear, the derivative
acting on $\phi''$ here gives what is usually called the Born correction, while the derivative
acting on the $\phi'$ term inside the inner integral gives the lens-lens correction.
We can rewrite the perpendicular deflection as
\be
x_\perp^i(\vec\theta) = -2 \int_0^{\chi_s} d\chi (\chi_s-\chi) 
{\partial \over\partial x^i}  
\phi\big(\vec x(\chi,\vec\theta)\big) 
\eql{final}\ee
where $\phi$ and $\vec x$ now include all orders in perturbations:
$\phi=\phi^{(1)} + (1/2)\phi^{(2)} +\ldots$ and similarly for $\vec x$. 
Eq.~\Ec{final} is the usual starting point for ray-tracing simulations. We have
now rigorously justified the standard ray-tracing approach, at least to second
order in geodesic corrections.

\section{Conclusions}

We have jusitifed the standard ray tracing formula of gravitational
lensing. This argument connects much of the formal work on second order
perturbations in cosmology with the more phenomenological approach used
in lensing studies. A similar result holds for the luminosity 
distance: the first order expression for the luminosity
distance~\cite{pyne} can be used as the starting point for the second order
result, without going back to the geodesic equation.

Our primary result, \ec{final}, holds only on sub-horizon scales. On very large
scales, many of the corrections we were able to neglect are of the same order as
the Born correction and the nonlinear evolution of the potential. However, on
large scales, we expect all such corrections to be small, of order $10^{-5}$,
since they are not enhanced by spatial variations. Therefore, our results are
valid in the regime in which corrections are measurable. The one possible caveat
to this claim is if super-horizon perturbations can influence the local
observables\cite{kmnr}.

This work is supported by the DOE and by NASA grant NAG5-10842.

\end{document}